\documentclass[10pt,journal,compsoc]{IEEEtran}
\usepackage{hyperref}
\usepackage[utf8]{inputenc}
\usepackage[T1]{fontenc}
\usepackage{graphicx}
\usepackage{subfigure}
\usepackage{float}
\usepackage{stfloats}
\usepackage{amssymb}
\usepackage{amsmath}
\usepackage{color}
\usepackage{bm} 

\newcommand{\smallotimes}{\mathbin{\mathpalette\make@small\otimes}}
\newcommand{\change}[1]{{\color{black}{#1}}}

\usepackage{tikz,xcolor,hyperref}
\definecolor{lime}{HTML}{A6CE39}
\DeclareRobustCommand{\orcidicon}{%
    \begin{tikzpicture}
    \draw[lime, fill=lime] (0,0) 
    circle [radius=0.16] 
    node[white] {{\fontfamily{qag}\selectfont \tiny ID}};    \draw[white, fill=white] (-0.0625,0.095) 
    circle [radius=0.007];    \end{tikzpicture}
    \hspace{-2mm}}
\foreach \x in {A, ..., Z}{%
    \expandafter\xdef\csname orcid\x\endcsname{\noexpand\href{https://orcid.org/\csname orcidauthor\x\endcsname}{\noexpand\orcidicon}}
    }

\begin{document}
%
\title{Improving link prediction accuracy of network embedding algorithms via rich node attribute information}

\author{Weiwei Gu,
        Jinqiang Hou,
        Weiyi Gu
\IEEEcompsocitemizethanks{\IEEEcompsocthanksitem ~Weiwei Gu and Jinqing Hou, UrbanNet Lab, College of Information Science and Technology, Beijing University of Chemical Technology, Beijing 100029, China.\protect\ E-mail:weiweigu@mail.buct.edu.cn; 2022200847@buct.edu.cn

\IEEEcompsocthanksitem ~Weiyi Gu, School of Computing, China University of Geosciences, No. 388 Lumo Road, Wuhan, P.R. China. \protect\ E-mail:weiyigu2002@sina.com, Weiyi Gu is the corresponding author.}
}

\markboth{Journal of Social Computing}%
{Shell \MakeLowercase{\textit{et al.}}: Bare Advanced Demo of IEEEtran.cls for IEEE Computer Society Journals}

\IEEEtitleabstractindextext{%
\begin{abstract}
Complex networks are widely used to represent an abundance of real-world relations ranging from social networks to brain networks. Inferring missing links or predicting future ones based on the currently observed network is known as the link prediction task. Recent network embedding based link prediction algorithms have demonstrated ground-breaking performance on link prediction accuracy. Those algorithms usually apply node attributes as the initial feature input to accelerate the convergence speed during the training process. However, they do not take full advantage of node feature information. In this paper,besides applying feature attributes as the initial input, we make better utilization of node attribute information by building attributable networks and plugging attributable networks into some typical link
prediction algorithms and naming this algorithm Attributive Graph Enhanced Embedding (AGEE). AGEE is able to automatically learn the weighting trades-off between the structure and the attributive networks.
Numerical experiments show that AGEE can improve the link prediction accuracy by around 3\% compared with link  prediction framework SEAL, Variational Graph AutoEncoder (VGAE), and Node2vec.
\end{abstract}

}

\maketitle

\IEEEpeerreviewmaketitle

\ifCLASSOPTIONcompsoc
\IEEEraisesectionheading{\section{Introduction}\label{sec:introduction}}
\else
\section{Introduction}
\label{sec:introduction}
\fi

\IEEEPARstart {n}{etworks} (a.k.a. graphs) consist of entities (nodes) and their connections (links) which are a fundamental representation of many real-world relations ~\cite{barabasi2002linked}. For example, networks can be used to describe the Protein-Protein interaction in biology~\cite{theocharidis2009network}, the syndication investment events between venture capital institutions in economics ~\cite{gu2019exploring,ruiqi2023capital,qingyaosyndication}, and the structural or functional interaction between different brain regions~\cite{de2017multilayer}. For WWW, social networks, and citation networks, link prediction can also help in recommending relevant pages, finding new friends, or discovering new citations \cite{craven2000learning,popescul2003statistical,liben2007link}. These linkages between entities contain rich information on node properties, network structures, and network evolution. Predicting the existence of a relation, which is always abbreviated as link prediction, is a crucial task in network science not only in theory but also in practice. For networks in biology like protein-protein interaction networks, metabolic networks, and food webs, the discovery and validation of links require significant experimental effort. Instead of blindly checking all possible links, link prediction can help scientists to focus on the most likely links, which can sharply reduce the experimental cost. 

The conventional link prediction methods can be divided into several groups. The approaches that make link prediction according to local similarity are based on the assumption that two nodes are more likely to be connected if they have many common neighbors \cite{barabasi1999emergence,zhou2009predicting}. These approaches are fast and highly parallel since they only consider local structure. However, the biggest drawback is their low prediction accuracy, especially when the network is sparse and large. While, global similarity-based methods use the whole network topological information to calculate the similarity between links \cite{liu2013hidden,zhou2009predicting,rucker2012network,shang2022local}. Although those methods perform better on prediction accuracy, they usually suffer from high computational complexity problem which makes them unfeasible for graphs that contain million and billion of nodes. There are also some probabilistic and statistical-based approaches, assuming that there is a known prior structure of the network, like a hierarchical or circle structure. However, they can not get over the problem of low accuracy. Furthermore, the conventional approaches can hardly reveal hidden information about node properties and network structures behind the linkages. 

Recently, there has been a surge of algorithms that seek to make link predictions through network embedding which extracts both local and global structural information about nodes from graphs automatically. The idea behind these network embedding algorithms is to learn a mapping function that embeds nodes as points in a low-dimensional space $\mathbb{R}^d$ which encodes information from the original graph. Network embedding-based methods, which are usually based on the Skip-Gram method or matrix factorization, such as DeepWalk, node2vec, LINE, and struc2vec \cite{perozzi2014deepwalk, grover2016node2vec, ou2016asymmetric, tang2015line, ribeiro2017struc2vec}, have achieved a much higher link prediction accuracy compared with the conventional ones. Random walk-based network representation learning algorithms are task agnostic, and the learned representations are used to perform graph-based downstream machine learning tasks, such as node classification~\cite{kipf2016semi}, node centrality measuring~\cite{gu2021discovering} , as well as link prediction \cite{gu2017hidden,grover2016node2vec,kipf2016variational}. To start with, there is no supervised information during the training process, the representation of nodes is updated directly without considering the global structure of networks. Besides , computational complexity is the biggest bottleneck since Skip-Gram-based algorithms require a large number of random walks \cite{perozzi2014deepwalk, grover2016node2vec,ribeiro2017struc2vec}, and they have limited expressive power since the embedding process is fixed by random walk strategies. Besides, the representations can be hardly extended for inductive learning since the embedding vectors can not be transferred from graph to graph\cite{hamilton2017inductive}.

\begin{figure*}[t]
\begin{minipage}[t]{0.45\linewidth}
\centering
\includegraphics[scale=0.5]{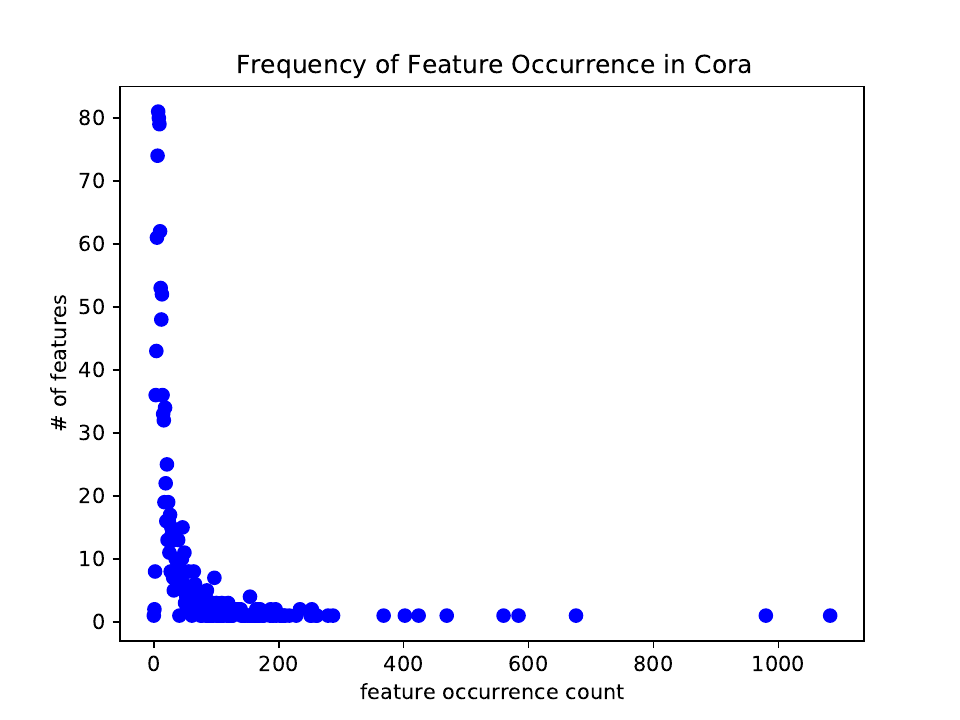}
\end{minipage} 
\begin{minipage}[t]{0.45\linewidth} 
\centering
\includegraphics[scale=0.5]{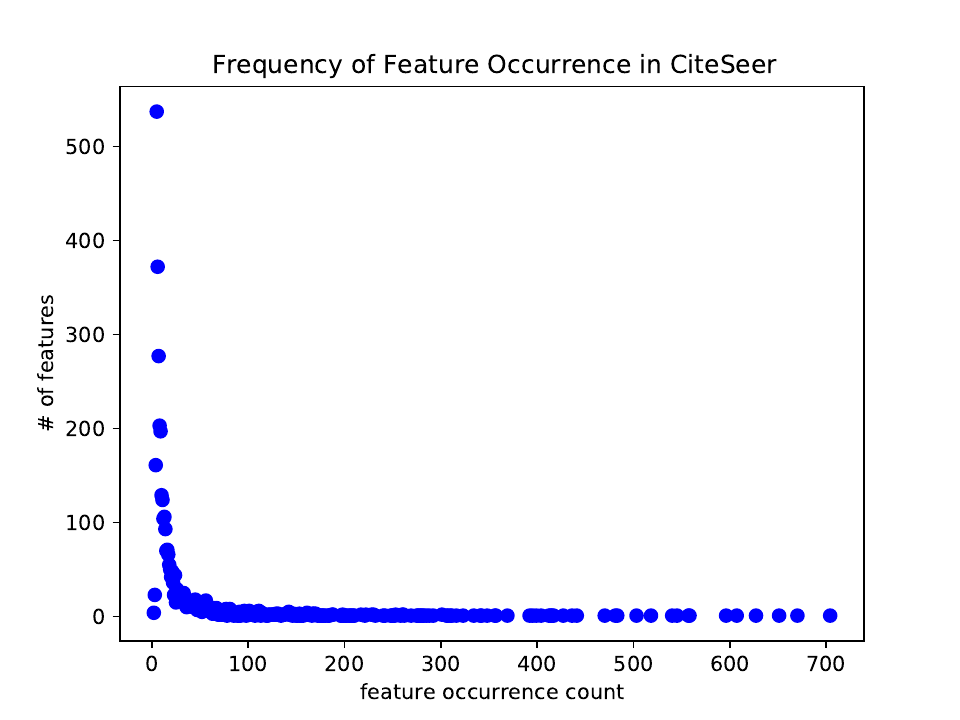}
\end{minipage}

\caption{\label{fig:cora_distribution} The attributes distribution of the Cora (left panel) and CiteSeer (right panel) feature matrix with a large number of attributes occurring occasionally, while a small number of attributes contribute as features of many nodes.}
\end{figure*}

\change{In recent years}, deep learning techniques based on graph neural networks have achieved triumphs in image processing \cite{he2016deep} and natural language processing \cite{gehring2016convolutional}. This stimulates the extensions of the methods on graph structures to perform link prediction tasks by converting the network structures into low-dimensional representations. Those algorithms borrow the concept of convolution from the convolutional neural network and convolve the graph directly according to the connectivity structure of the graph via the Graph Neural Networks (GNNs) architecture. Representative algorithms in this genre include Graph AutoEncoder \cite{kipf2016variational}, GraphSage \cite{hamilton2017inductive}, and SEAL \cite{zhang2018link}. Compared with traditional link prediction methods, \change{GNN-based link prediction algorithms take advantage of the nodal attributes via initialized node representations with feature matrix to accelerate the convergence speed during the optimization process. Each row of the feature matrix corresponds to a node's attributes in a graph, with one column representing a feature of all nodes. 

To the best of our knowledge, all the GNN-based link prediction algorithms only utilize node attributes as the initial feature input and ignore the rich nodal information contained in the feature matrix. For example, in machine learning articles, ``deep learning'' is a widely used keyword in an abundance of articles, and due to its extensive use in many articles, it has a limited contribution to the linkages prediction of the citation network. However, the co-occurrence of rare keywords such as ``percolation'' in machine learning articles may reveal some significant characters since this keyword generally only appears in some multidisciplinary papers. Feature frequency is a crucial indicator for predicting article citation, especially for finding the connection between interdisciplinary and innovative papers.

Defining and identifying feature information contained in the feature matrix provides a way to discover latent graph connections that can not be quantified by network topology. } 
Figure~\ref{fig:cora_distribution} shows the feature co-occurrence of the Cora\change{\cite{mccallum2000automating}} and CiteSeer \change{\cite{bandyopadhyay2005link}}networks. From Fig.~\ref{fig:cora_distribution} we find the co-occurrences of node features follow a power law distribution with some features widely existing in most of the nodes. For example, the most frequent feature in the Cora graph appeared 1,083 times, which means in the Cora graph, nearly 40\% of nodes take this feature as a keyword. We also discovered some keywords (attributes in the feature matrix) rarely occur, they exist only in several nodes, and the rarely appeared features contain rich information about nodal link information. \change{How to quantify feature differences and better utilize feature information to enhance link prediction accuracy is an important link prediction research field.} 

In this paper, in order to make better utilization of the feature attributes to improve link prediction accuracy, we propose Attributive Graph Enhanced Embedding (AGEE). 
AGEE is a plug-and-play algorithm, it can be plugged into a bunch of link prediction algorithms without modifying the original link prediction architecture, and it also has the universal property, which means most of the link prediction algorithms can enhance their prediction power with the help of AGEE. \change{AGEE first uses entropy to quantify information of each attribute and then computes the total amount of information between any node pairs. After that, AGEE sets a threshold to the total amount of information between nodes to build a feature graph. In the last step, it separately trains the feature graph and structure graph with different training algorithms to find a trade-off between the feature graph predicted probability and the structure graph to form the final structure link prediction accuracy. We validate AGEE's performance on node2vec, Variational Graph AutoEncoder (VGAE), and SEAL algorithms over several networks, and the results show that AGEE can significantly improve link prediction accuracy by around 3\%.}

\section {Methods}
\change{AGEE consists of two parts, building feature graphs according to the given feature matrices and plugging the predicted results into a variety of algorithms to improve the link prediction accuracy.}

\subsection{Build Feature Graph}

Recent graph neural network-based algorithms have achieved the state of the art accuracy in link prediction tasks. However, those algorithms only take node feature information as the initial input of their models and ignore the rich hidden information behind them. The feature matrix of a graph is denoted as \textbf{F}, which is an $N \times M$ matrix, with $N$ representing node number and $M$ representing feature number. For example, in Fig.~\ref{fig:cora_distribution}, in the Cora dataset each node has 1,433 features, three of which appeared in over half of the nodes, in other words, those features are widely used as keywords, and it is difficult to infer the link existence probability between node pairs based on the information provided by those attributes. We also notice that there are a large number of rare features, which co-exist only in several nodes, and those rare features may represent field-specific and vital keywords in sub-field research areas, and the connection probability between those nodes based on those rare keywords is higher compared with the widely appeared features. Feature frequency is an important indicator to quantify features' importance and tie strength between nodes. The co-occurrence of less frequent features indicates a tight relation while the commonly appeared features represent a loose relation.

\change{In order to distinguish attributes and quantify information contained within different features, we proposed the following Equation~\ref{eq:self-information}
} 

\begin{equation}
\centering
 \label{eq:self-information}
   \bm{I(m)} = -{\log_{2}(p(m)) }
\end{equation}
where $p(m)$ is the frequency that feature $m$ appears over the total number of features across all nodes. $\bm{I}(m)$ quantifies information convery by feature $m$. Note that the base of the logarithm only affects the scaling, and here we take base 2 in units of bits to quantify feature information.

\begin{figure*}[bp]
\begin{minipage}[bp]{0.45\linewidth}
\centering
\includegraphics[scale=0.5]{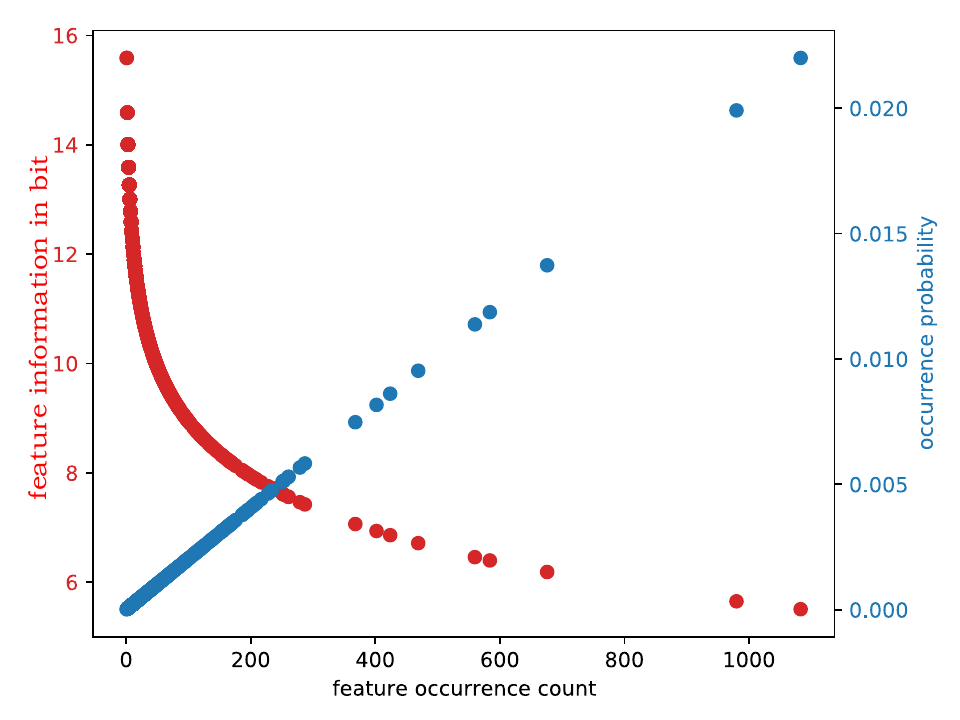}
\end{minipage}
\begin{minipage}[bp]{0.45\linewidth}
\centering
\includegraphics[scale=0.5]{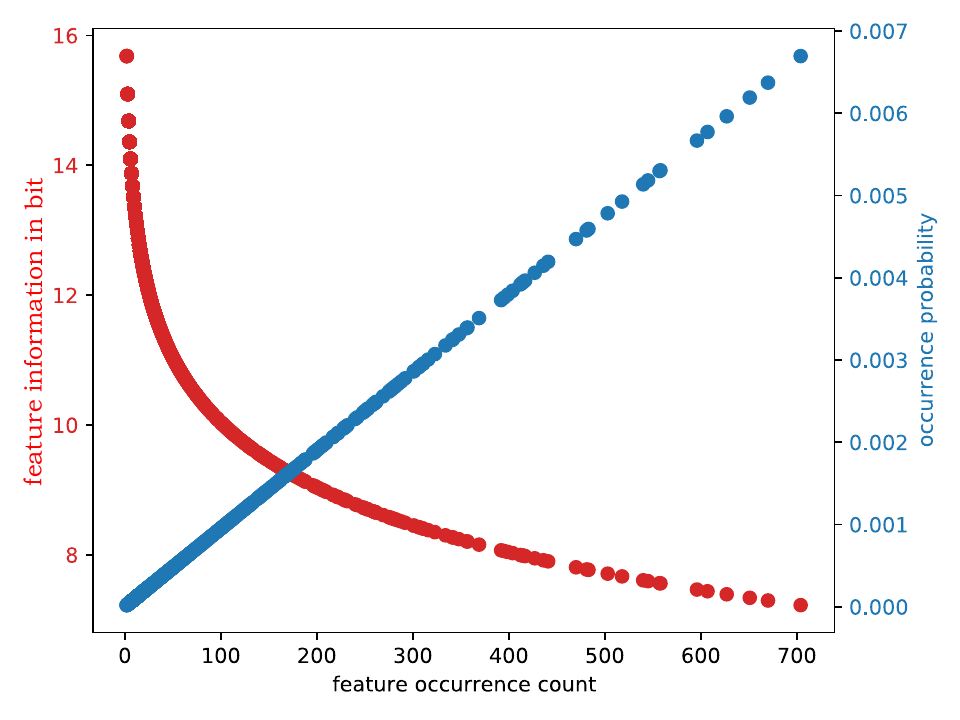}
\end{minipage}%

\caption{\label{fig:cora-inforamtion_amount} Comparison of \change{feature}-information quantified by eq.\ref{eq:self-information} and feature occurrence probability of Cora (left panel) and CiteSeer (right panel) datasets.}
\end{figure*}

\change{Feature} information is also called the ``surprisal'' of a feature, with low occurrence probabilities corresponding to higher information, and frequently appearing features carry a small amount of information. We concatenate \change{feature}-information of feature into a feature information vector $\bm{I}$ = \{$\bm{I}(1)$, $\bm{I}(2)$, . . . , $\bm{I}(M)$ \}, $\bm{I}$ $\in$ $\mathbb{R}^M$. As shown in Figure~\ref{fig:cora-inforamtion_amount}, the left vertical axis represents the \change{feature} information bits calculated with Eq.\ref{eq:self-information}, and the right vertical axis represents the frequent probability over all node features. \change{As shown in Fig.\ref{fig:cora-inforamtion_amount}, Eq.\ref{eq:self-information} gives high information to less frequently occurred features and low information to the frequently appeared features, especially for nodes that appeared greater than 1000.}

\change{Equation~\ref{eq:self-information} quantifies information contained in each feature. We propose Eq.\ref{eq:entropy_matrix} to measure the feature relations between nodes.}

\begin{equation}
\centering
 \label{eq:entropy_matrix}
 \textbf{W} = (\bm{I}\times \textbf{F} ) \cdot (\bm{I}\times \textbf{F} )^{T}
\end{equation}

where $\bm{I}$ is a $1 \times M$ dimensional vector with its component representing the feature information, \textbf{F} denotes the original feature matrix, and ($\bm{I} \times \textbf{F}$) $\in$ $\mathbb{R}^{N \times M}$ represents weighted feature matrix, with each element not only indicating the existence of the feature but also quantifying the feature's influence on node. In Eq.\ref{eq:entropy_matrix} we use $\textbf{W}$ $\in$ $\mathbb{R}^{N \times N}$ to represent feature relations between the node pairs with each element $\textbf{W}{i,j}$ stands for the feature similarity strength between nodes $i$ and $j$. 

\change{Matrix \textbf{W} is densely connected with each element representing feature similarity which is computed based on the cumulative information that features contain. The higher the feature similarity, the easier it is for nodes to connect together \cite{papadopoulos2012popularity}.} Based on the weighted similarity matrix \textbf{W} we build the feature network $G_{feature} =(V, E)$. $G$ is an undirected and unweighted graph, where $V$ denotes the set of vertices $|V| = N$ which is consistent with nodes in the structure graph $G\_structure$, $E \in V \times V$ is the set of feature links. The adjacency matrix of the feature graph is denoted as matrix \textbf{A}, where $\textbf{A}_{i,j}$ = 1 if (i,j) $\in E$ otherwise $\textbf{A}_{i,j}$ = 0 if there is no feature link between $i$ and $j$. 

\change{In this paper, the feature edges are generated with the following process. We first set a benchmark value $\epsilon$ to binarize the feature weight matrix \textbf{W}, and we set that $\textbf{A}_{i,j}$ equals to 1 if $\textbf{W}_{i,j}$ is greater than $\epsilon$, otherwise $\textbf{A}_{i,j}$ equals to 0, as shown in Eq.~\ref{eq:adj}. Most networks have scale-free properties, and small $\epsilon$ values lead to dense networks while large $\epsilon$ values lead to sparse networks. In order to mimic the real density of the graph, we set $\epsilon$ to the standard value which features network density $\textbf{A}_{i,j}$
equal to the precomputed structure network density.}

\begin{equation}
\centering
\label{eq:adj}
\textbf{A}_{i,j}  = \begin{cases} 1 & { \textbf{W}_{i,j} > \epsilon} \\ 0 &  { \textbf{W}_{i,j} <= \epsilon} \end{cases}
\end{equation}

\subsection{Plug AGEE into Link Prediction Algorithms}

\change{AGEE is a plugin that can be embedded into other link prediction algorithms to increase their link prediction accuracy.} In this paper, we select \change{three typical and widely used link prediction algorithms with network embedding fashions, which are} node2vec\cite{grover2016node2vec}. VGAE~\cite{kipf2016variational}, and SEAL~\cite{zhang2018link}. Node2vec mainly applies the skipGram~\cite{mikolov2013efficient} which is an unsupervised learning technique, VGAE is mainly based on variational auto encoders~\cite{kingma2019introduction} which is a probabilistic generative model that requires neural networks as a part of the overall structure. SEAL uses node classificational Networks~\cite{kipf2016semi} which applies an efficient variant of convolutional neural networks that operate directly on graphs. \change{In this section, we take the node2vec algorithm as an example and show how the AGEE helps improve its link prediction accuracy. A brief introduction to VGAE and SEAL is shown in the experiments section.}

\subsubsection{Learning Node Representations}
Node2vec is one of the most popular network embedding and link prediction algorithms, and is widely used to solve a variety of theoretical and practical problems in biology~\cite{kim2018relation}, neurosciences~\cite{rosenthal2018mapping} and social science~\cite{gu2017hidden}. Node2vec builds on the word2vec algorithm~\cite{mikolov2013distributed} by 
comparing nodes in the network to ``words'' and a sequence of nodes explored during a biased random walk to ``sentence''. After the analogy process, the generated ``node sentences'' are fed into the skip-gram model to get the feature representation of nodes. 

\change{Next, we will introduce node2vec and illustrate how it and the mapping function $f: \bm{h} 
\leftarrow \mathbb{R}^d$ learns representation of the feature graph $G\_feature$. Here $d$ is a parameter specifying the number of dimensions, and function $f$ is a matrix that contains $\bm{|h|}\times d$ parameters. For each node $i$ in the feature graph, $N_S(i)\subset \bm{h}$ are a subset of network neighbors of node $u$. Node2vec applies the Skip-Gram architecture to optimize the following objective function which maximizes the log-probability of the observing nodes given its neighborhood  $N_S(i)$ and mapping function $f$: 
\begin{equation}
\max _f \sum_{i \in V} \log \operatorname{Pr}\left( N_S(i) \mid f(i)\right) .
\end{equation}  }

\change{To make the optimization problem manageable, we adopt the following two common assumptions:
We first assume that the likelihood of observing a neighborhood node is independent of observing any other neighborhoods giving the representation of the source node $i$.
\begin{equation}
\operatorname{Pr}\left(N_S(i) \mid f(i)\right)=\prod_{n_j \in N_S(i)} \operatorname{Pr}\left(n_j \mid f(i)\right)
\end{equation}

Influence symmetric is another assumption that assumes each node pair has a symmetric effect in the representation space. The conditional likelihood of source-neighborhood node pair can be parametrized by a dot product of their representations:
\begin{equation}
\operatorname{Pr}\left(n_j \mid f(i)\right)=\frac{\exp \left(f\left(n_j\right) \cdot f(i)\right)}{\sum_{v \in V} \exp (f(v) \cdot f(i))} .
\end{equation}

With the above assumptions, the objective function can be written as

\begin{equation}
\label{eq:node2vec_optimize}
\max _f \sum_{i \in V}\left[-\log Z_i+\sum_{n_j \in N_S(i)} f\left(n_j\right) \cdot f(i)\right]
\end{equation}
where the per-node partition function,$Z_i=\sum_{j \in V} \exp (f(i) \cdot f(j))$, is expensive to compute for large networks and we approximate it using negative sampling~\cite{mikolov2013distributed}, node2vec optimize eq.~\ref{eq:node2vec_optimize} with stochastic gradient ascent over the model parameters defined on representation learning function $f$. In this paper, we learn the node2vec representation of the feature graph and structure graph with the implementation from {node2vec}\href{https://github.com/aditya-grover/node2vec}. }

\change{\subsubsection{Links Existence Probability}}

\change{ $\bm{h}_{feature}^{i}$ represents the learned node embedding of node $i$ of feature graph $G\_{feature}$, for every pair of nodes $i$ and $j$ in $G\_{feature}$ or $G\_{structure}$, we use the Hadamard product to represent the potential relations of node pairs. 
${\bm{e}_{feature}^{i,j}}$ =$\bm{h}_{feature}^{i}$ $\otimes$ $\bm{h}_{feature}^{j}$, Note that $\bm{e}_{feature}^{i,j}$ is the relation representation between node $i$ and $j$ which is also a d-dimensional vector. The above procedure is also applied to the structure graph. Link prediction between node $i$ and $j$ in the feature graph can be represented as:}
\begin{equation}
\label{Logistic}
  p_{i,j}(\bm{e}_{feature}^{i,j};\bm{\theta}) =  \frac{1}{1+\exp{(\bm{e}_{feature}^{i,j}} \bm{\theta})}
\end{equation}
where $\bm{\theta}$ is a $d$-dimensional parameter vector, and ${\bm{e}_{feature}^{i,j} \bm{\theta}}$ is the dot product between the vectors $\bm{e}_{feature}^{i,j}$ and $\bm{\theta}$. The best estimate of the entries of vector $\bm{\theta}$ is obtained from the training set via logistic regression.

In this paper, we first hide $\gamma$ randomly chosen edges of the structure graph to form the “positive” sample set. We then sample an equal amount of disconnected vertex pairs as the “negative” sample set. The union of the “positive” and “negative” samples formed the test set and are denoted as $S_{structure}$ of the structure graph. We select 10\% percent of the remaining edges and an equal amount of disconnected node pairs to form the validation set. The rest edges and an equal amount of disconnected node pairs form the training set. We focused our task on predicting the existence or absence of the relations in $S_{structure}$.

\begin{figure*}[t]
\centering
\includegraphics[scale=0.5]{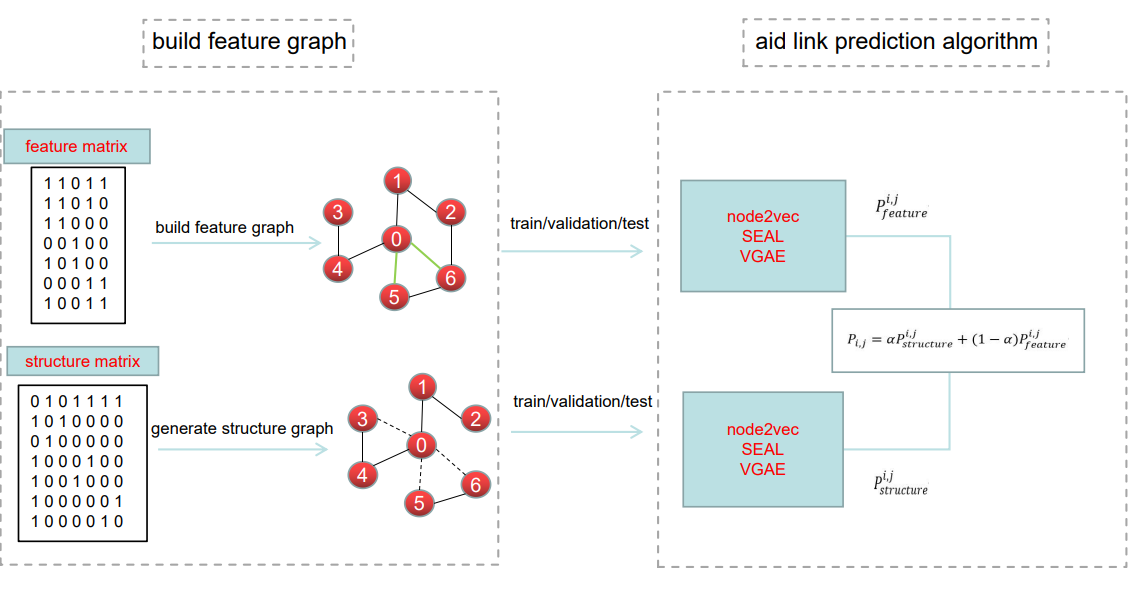}
\caption{\label{fig: illustration}The overall architecture of AGEE algorithm. We use a simple seven-node graph with five nodal features to illustrate AGEE’s architecture. In the first stage, based on the feature matrix we build the feature graph according to Eq.~\ref{eq:self-information}, ~\ref{eq:entropy_matrix} and ~\ref{eq:adj}, we also build the structure graph according to the adjacency matrix, in the structure graph, we use the dashed lines to indicate some of the representative edges that need to be predicted by link prediction algorithms and apply solid lines to represent edges used to train the link prediction algorithms. In the structure graph, there are no edges between nodes 0 and 6, 5 and 0, in the feature graph, there are links between them. The featured graph works as a supplementary for the structure graph. 
In the second stage, we train the feature graph and structure graph separately and use a hyper-parameter $\alpha$ to find the trade-off between the structure link prediction probability and the feature link prediction probability. }
\end{figure*}

\change{\subsubsection{Plug Feature Predicted Probability into Structure Graph}}

we train the feature graph and structure graph and plug feature predicted probability into the structure graph with a hyper-parameter $\alpha$ which balances the importance between the two graphs. Eq.~\ref{eq:alpha} is the mathematical aggregation function. In order to compute the edge existence probability between $i$ and $j$ of a structure graph, for example, nodes 0 and 6 in Fig.~\ref{fig: illustration}, $\alpha$ acts as the consensus coefficient between the predicted probability of the original structure graph ${p}_{structure}^{i,j}$ and feature graph ${p}_{feature}^{i,j}$. Our intuition behind the consensus operation lies in that the link between nodes 0 and 6 is difficult to predict in the original structure graph due to the absence of common neighbors. However, since in the feature graph nodes 0 and 6 have a common neighbor 5, the feature graph will give a high link probability between 0 and 6. Our experimental results show that the optimal value of $\alpha$ is around 0.6, which represents that node pairs in the test set of the structure graph consider 0.6 of their own predicted probability from the original structure graph and 0.4 from the feature graph. 

The overall illustration of the aggregation process is shown in Fig.~\ref{fig: illustration}. AGEE algorithm consists of two stages. In the first stage, we build a feature graph according to rich nodal attributes and feature entropy. In the second stage, we separately train the feature graph and the original structure graph since they follow different dynamics, and we introduce a hyper-parameter $\alpha$ that trades off the feature and structure prediction probabilities.

\begin{equation}
\centering
\label{eq:alpha}
{p}_{i,j}  = \alpha {p}_{structure}^{i,j}+(1-\alpha){p}_{feature}^{i,j}
\end{equation}

\section {Experiments}
\subsection{Datasets, Baselines and Evaluation Metrics}

We use three citation datasets with rich node attribute information. The Cora dataset consists of 2708 scientific publications classified into seven classes and 5429 links with each publication in the dataset being described by a 0/1-valued word vector indicating the absence/presence of the corresponding word from the dictionary. The dictionary consists of 1433 unique words as nodal attributes. The CiteSeer dataset consists of 3312 scientific publications classified into one of six classes. CiteSeer has 4732 links and 3703 nodal attributes by 0/1 values that respond to the absence/presence of the corresponding word from the dictionary. The PubMed is a larger dataset compared with Cora and CiteSeer and it contains 19717 scientific publications from the PubMed database pertaining to diabetes classified into three classes. It has 44338 links with each publication in the dataset described by a Term Frequency-Inverse Document Frequency (TF/IDF) weighted word vector from a dictionary which consists of 500 unique words.

\change{Variational Graph Autoencoder (VGAE) is a framework for unsupervised learning on graph-structured data based on the Variational AutoEncoder (VAE) architecture. VGAE improves link prediction accuracy on a number of network datasets such as Cora and CiteSeer, and it has inspired a wide range of ongoing research. VGAE is also the most promising model for network constriction tasks. The main idea of VGAE is that it encodes the input into a distribution rather than a high-dimensional vector space. Then a random sample is taken from the distribution rather than generated from the encoder directly. The loss function of VGAE consists of two parts. The first part is the variational lower bound, which measures how well the output network reconstructs the original graph. If the reconstructed network is dissimilar from the original data, then the reconstruction loss will be high. The second part works as a regularization. It is the Kullback–Leibler (KL) divergence of the approximate from the true posterior, which measures how closely the output distribution matches the latent network distribution. In VGAE, we use the default parameter setting and the PyTorch implementation from \href{https://github.com/Flawless1202/VGAE_pyG}{VGAE\_PyTorch}.

SEAL is a novel link prediction framework and has achieved state-of-the-art link prediction accuracy in a large number of small-scale networks. Instead of applying the entire network information to do the link prediction task, SEAL first extracts local enclosing subgraphs within a 2-hops neighborhood. In doing so, SEAL enables graph feature learning ability, and through node labeling operation, SEAL can better capture the hidden relationship of the nodes in subgraphs. After the subgraph extracting process, SEAL applies a Graph Neural Network (GNN) to replace the fully-connected neural network. During the graph convolutional process in the GNN framework, SEAL permits learning from not only subgraph structures but also latent and explicit node features, thus incorporating multiple types of information. SEAL outperforms the previous state-of-the-art method on link prediction tasks. It is the pioneer to apply node labeling and subgraph extracting operations to the link prediction task. SEAL treats link prediction as a subgraph binary classification task. It outperforms previous latent embedding-based link prediction algorithms such as VGAE and node2vec on small datasets. As for larger graphs that contain millions or billions of nodes, SEAL has a memory error problem even when the network is not big. We find SEAL fails to do the link prediction task even for networks that contain less than twenty thousand nodes such as the PubMed network. In SEAL we apply the PyTorch implementation from \href{https://github.com/muhanzhang/SEAL}{SEAL\_PyTorch}.}

In this paper, we use The area under the ROC curve (AUC) as an evaluation metric to measure the performance of the link prediction application. AUC is computed for every test node and the average values are reported. A higher AUC indicates better predictive performance. In this paper, for every link prediction accuracy, we repeat the link prediction procedure 10 times and get the mean average results over these repetitions.

\subsection{Link Prediction Accuracy Comparison}
 AGEE algorithm encodes structure network information as well as feature network information by optimizing their network topology separately and learning the best trade-off between the predicted probability of structure prediction and feature prediction. We plug AGEE into node2vec, VGAE, and SEAL algorithms.Although the main ideas and architectures behind those algorithms are different, they all follow the similar training process described below. In this paper, to do the link prediction task, we apply the repeated random sub-sampling validation by applying the following procedure 10 times. We first hide 10\% of randomly chosen edges of the original structure graph. The hidden edges in the original graph are regarded as the “positive” sample set. We sample an equal amount of disconnected vertex pairs as the “negative” sample set. The union of the “positive” and “negative” sample sets form our test set. The validation set also contains 10\% of randomly chosen edges of the original graph excluding the test set. The training set consists of the remaining 80\% of connected node pairs and an equal number of randomly chosen disconnected node pairs. We use the training set to learn the probability of connection between pairs of nodes, use the validation set to identify the best training epochs and trade-off weights, and use the test set to evaluate the performance of different link prediction algorithms. Besides the original node2vec, VGAE, and SEAL algorithms, we add comparisons with the following well-known link prediction algorithms. 

LINE\cite{tang2015line} minimizes a loss function to learn embedding while preserving the first and the second-order neighbors' proximity among vertices in the graph. GANR\cite{gu2021discovering} applies the node centrality network architecture to do the link prediction task and reveal the hidden structures of networks. GraphSAGE\cite{hamilton2017inductive} learns node embedding through a general inductive framework consisting of several feature aggregators. It usually adopts a supervised node classification task as the evaluation benchmark with the assumption that a better embedding algorithm leads to higher node classification accuracy. ARVGA~\cite{pan2018adversarially} uses a variational graph autoencoder to learn to embed and perform the link prediction as the supervised task.

From Table~\ref{tab:accuracy} we find that general algorithms with deep graph neural networks that have auto-encoder and auto-decoder frameworks such as GANR, GraphSAGE-mean, and ARVGA have higher AUC values compared with shallow neural network models such as LINE. When plugging our glue AGEE algorithm into node2vec, SEAL, and VGAE, the link prediction accuracy of all the previous algorithms improved by around 3\%. We also notice that even though node2vec has a shallow neural network, its performance is only slightly behind the deep graph neural networks-based algorithms, and the combination of AGEE and node2vec algorithm outperforms all link prediction algorithms and achieves the state-of-the-art link prediction accuracy of all datasets.  
\begin{table*}[t]
\centering
\caption{The comparison of AGEE enhanced algorithms over link prediction task under AUC metric score.}
\label{tab:accuracy}
\setlength{\tabcolsep}{2pt} 
{\strut\fontsize{9}{11}\selectfont 
\begin{tabular}{cccccccc|ccc}
\hline
AUC             & GANR    &LINE  & GraphSAGE-mean & ARVGA & node2vec & VGAE & SEAL &  AGEE\_VGAE & AGEE\_SEAL& AGEE\_node2vec \\ \hline

Cora & 0.93   & 0.76  & 0.89 &0.93 & 0.92 &0.91 & 0.90  & 0.93 & 0.93 & \textbf{0.95}  \\ 
CiteSeer   &0.91  &0.73 &0.90 &0.94 & 0.89 &0.90 & 0.90  &0.92 &0.93    & \textbf{0.96}      \\ 
PubMed  & 0.95 &  0.84& 0.91 & 0.95 &0.94 &0.93  & 0.94  & 0.96 & 0.95&  \textbf{0.97} \\
 \hline
\end{tabular}
}
\end{table*}

\subsection{Parameter Sensitivity Study}
\subsubsection{Link prediction Accuracy over Different Training Sets and Robustness Test}

\begin{figure*}[t]
\begin{minipage}[t]{0.45\linewidth}
\centering
\includegraphics[scale= 0.5]{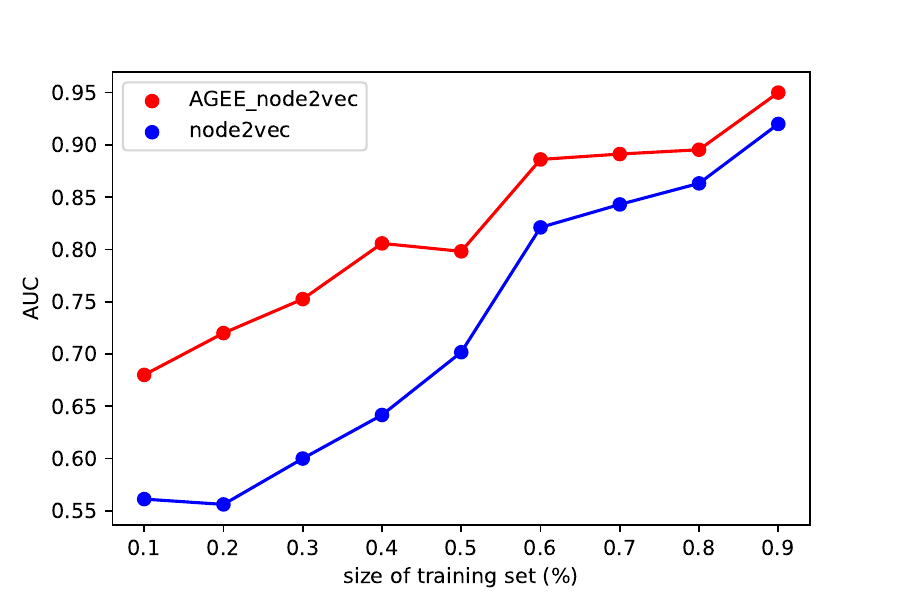}
\end{minipage}
\begin{minipage}[t]{0.45\linewidth}
\centering
\includegraphics[scale=0.5]{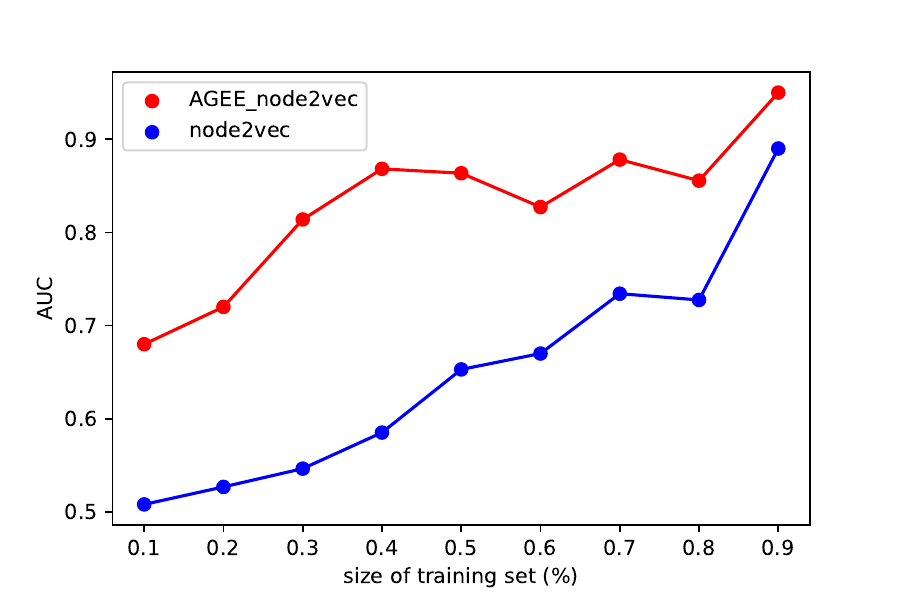}
\end{minipage}%

\caption{\label{fig:cora-citeseer-size} AUC score of link prediction accuracy comparison between original node2vec algorithm and AGEE\_node2vec on a variety of training sets for Cora(left panel) and CiteSeer (right panel).}
\end{figure*}

We quantify AGEE's link prediction ability by tuning the size of the training set. As shown in Fig. \ref{fig:cora-citeseer-size}, we find that AGEE\_node2vec outperforms the original node2vec algorithm especially when the training set is small. When we use only 10\% of the edges in the original network to perform the link prediction task, the node2vec's link prediction accuracy is 55\% only 5\% higher than random guess 50\%, however, for AGEE\_node2vec the accuracy is around 68\%. This result may due to that when the network structure is sparse, nodal feature network works as a complementary to provide rich link information about nodes. This experiment shows AGEE's ability to accurately predict node relations when the training set is small, compared with original link prediction algorithms.

Being able to predict missing links with a small fraction training set serves at least in the following two aspects. It first reveals the algorithm's ability to capture the hidden structure and the link principle of the network. Good link prediction algorithms can reveal the link principles among nodes even when the training set is small, and they have excellent inductive learning ability. Second, for some biological networks, such as brain connection networks, and protein-protein interaction networks, identifying links between nodes is expensive and sometimes involves bias, accurately predicting missing links or future ones without and experiments saves cost. 

\subsubsection{Link Prediction Accuracy Across Different Consensus Values}
\begin{figure}
\centering
\includegraphics[scale=0.50]{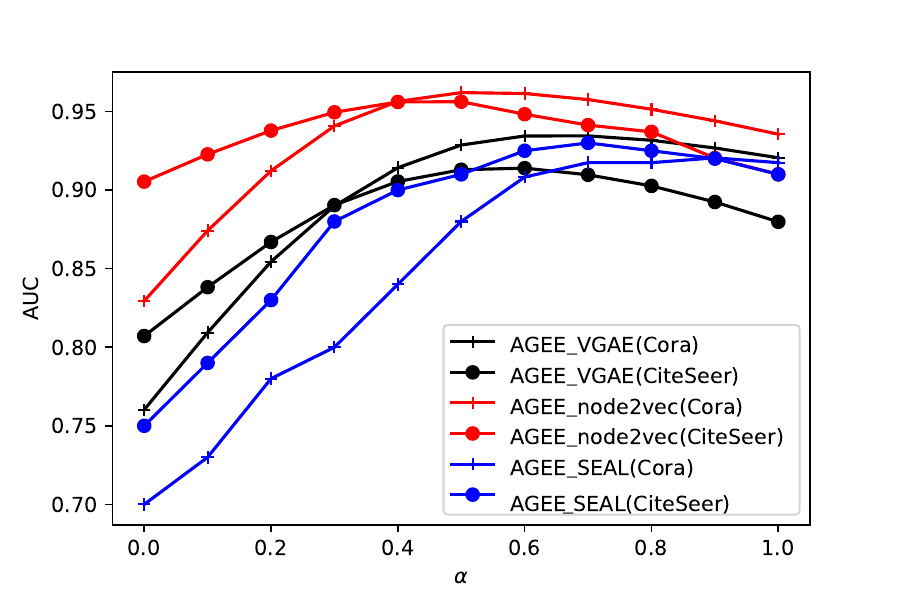}

\caption{\label{fig:turn_alpha} AUC score of link prediction accuracy across different prediction algorithms on Cora and CiteSeer network over hyper-parameter $\alpha$.}
\end{figure}

The consensus value $\alpha$ in  Eq.~\ref{eq:alpha} is one of the most important hyper-parameter in AGEE plugged algorithms. It determines the prediction probability trade-off between feature prediction and structure prediction. $\alpha$ equals 0 representing the condition in which AGEE plugged algorithms only take feature prediction accuracy to form the final link prediction accuracy, while $\alpha$ equals 1 representing it equals the original node2vec algorithms. In this part, we take AGEE\_node2vec as an example to test $\alpha$'s influence on link prediction outcomes. As shown in Fig. \ref{fig:turn_alpha}, we find $\alpha$ is quite robust over different datasets across different AGEE-combined link prediction algorithms. The optimal $\alpha$ value is around 0.6 which means when predicting structural link relations between nodes, we can take 60\% of the predicted probability from structures and 40\% of the probability from the nodal feature graphs. From Fig. \ref{fig:turn_alpha}, we also find among all the feature-enhanced link prediction algorithms, AGEE\_node2vec outperforms AGEE\_SEAL and AGEE\_VGAE that are famous link prediction algorithms and are based on graph convolutional network models.


\section*{Conclusion}
Node feature matrix usually works as the original input for graph neural networks to do the downstream classification or prediction tasks. In this paper, we extend the application of the node feature matrix by building a nodal feature graph, and with this nodal feature graph, we propose a plug-and-play $AGEE$ model that improves the link prediction accuracy of existing embedding-based link prediction algorithms without adding extra information and increasing the complexity of algorithms. Feature graph enhanced prediction algorithm can improve the link prediction accuracy by around 3\%. We introduce a trade-off hyper-parameter $\alpha$ to balance the importance between feature graph predicted probability and structure graph predicted probability. Our results find the consensus value $\alpha$ is quite robust, and $\alpha$ is usually around 0.6 which means when predicting structure links, the structure graph plays a more important role compared with the feature graph.   

Although the link prediction accuracy of AGEE\_node2vec is quite high compared with previous link prediction algorithms, there is still room for improvement. In AGEE\_node2vec the final link prediction probability is determined by a fixed $\alpha$ value, and all edges share the same consensus value. \change{However, link formation follows a popularity and similarity principle and shows that nodes tend to build links with more popular nodes, nodes that have similar features, or both. The node feature matrix is a good indicator to describe the similarity between nodes, and different links are supposed to put different emphasis on feature graphs and structure graphs. Further extensions of AGEE could involve assigning edges of each node pair a personalized $\alpha$ value to improve link prediction accuracy in a more general way. }


%





\ifCLASSOPTIONcaptionsoff
  \newpage
\fi

\end{document}